\documentclass[conference]{IEEEtran}
\IEEEoverridecommandlockouts
\usepackage{cite}
\usepackage{amsmath,amssymb,amsfonts}
\usepackage{algorithmic}
\usepackage{graphicx}
\usepackage{textcomp}
\usepackage{xcolor}
\def\BibTeX{{\rm B\kern-.05em{\sc i\kern-.025em b}\kern-.08em
    T\kern-.1667em\lower.7ex\hbox{E}\kern-.125emX}}
\begin{document}

\title{Mixed Reality: The Interface of the Future\\
}

\author{\IEEEauthorblockN{Dipesh Gyawali}
\IEEEauthorblockA{\textit{Department of Computer Science} \\
\textit{Louisiana State University}\\
Baton Rouge, Louisiana, United States \\
dgyawa1@lsu.edu}
}

\maketitle

\begin{abstract}
The world is slowly moving towards everything being simulated digitally and virtually. Mixed Reality (MR) is the amalgam of the real world with virtual stimuli. It has great prospects in the future in terms of various applications additionally with some challenges. This paper focuses on how Mixed Reality could be used in the future along with the challenges that could arise. Several application areas along with the potential benefits are studied in this research. Three research questions are proposed, analyzed, and concluded through the experiments. While the availability of MR devices could introduce a lot of potential, specific challenges need to be scrutinized by the developers and manufacturers. Overall, MR technology has a chance to enhance personalized, supportive, and interactive experiences for human lives.
\end{abstract}


\section{Introduction}
Technology is advancing at a rapid pace. The world is slowly moving towards everything being simulated digitally and virtually. Real-life problems are being solved by decision-making tools using advanced technology. It is significant to make the user interface easy to use for proper utilization of technology. To achieve this goal, many computing technologies have emerged such as Virtual Reality(VR), Augmented Reality(AR), and Mixed Reality(MR).

Generally, in today's digital age, there exists a real world, a virtual world, and a mix of both real and virtual. The real world is the physical world where everything exists physically and dynamically. The virtual world is a form of simulation where users can experience different contexts other than the physical world at the same time. The Hybrid world consists of a mixture of physical and virtual entities where anything can be manipulated over the physical world virtually. 

Mixed reality (MR) is a hybrid technology where the real world merges with virtual stimuli. The term 'Mixed Reality' was coined by Milgram and Kishino[1] in a paper "The Taxonomy of Mixed Reality Virtual Devices" in 1994. It includes AR and VR technologies that overlay virtual objects over physical entities. MR technology is the outcome of fusing the physical world. MR can be experienced using special devices e.g. Microsoft Hololens or applications of mobile phones which introduce the next potential sectors that change the world.

Mixed Reality is different from AR[4] and VR[2] in the context of how objects are overlayed and presented. Virtual Reality is sensed through VR devices that are mounted on the head where users can interact and mimic their natural behavior such as zooming in and zooming out the object, changing the angle of projection while moving, etc. There is no physical world in VR where users can interact with the entities physically. Augmented Reality is different from VR and can be achieved using AR devices or mobile phones. e.g. Snapchat application. In AR, there is an overlay of virtual objects over the physical entities where the interaction between user and object takes place. They can change different environment settings as well. Mixed Reality[3] is achieved using a typical MR computing device which establishes new settings of experiences by the interaction of virtual and physical space. Using MR, users themselves can manipulate the virtual entities in the physical world to achieve different configurations and environments.

In this research, various applications of Mixed Reality are studied, and found how these applications could shape the interface of the future. Additionally, there are some challenges that need to be considered while using MR technologies which are also presented in this research. The major contributions of this paper are:

\begin{itemize}
\item Study the various applications of Mixed Reality and its future prospects.
\item The way Mixed Reality could be the interface of the future.
\item The challenges faced when using Mixed Reality technology.
\end{itemize}

 \begin{figure}[hbt!]
  \centering
  \includegraphics[scale=0.50]{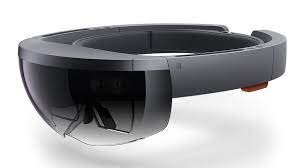}
  \caption{Microsoft Hololens}
  \end{figure}

\section{Literature Review}
Mixed Reality is an emerging technology where big tech companies and academic institutions are performing research extensively. Speicher et al. presented basic concepts about mixed reality and characterized MR applications in terms of the environment number, users, immersion and virtuality level, interaction degree, input, and output[3]. Similarly, the review of mixed reality, its current trend, challenges, and prospects have been studied by Rokhsaritalemi et al. (2020). They have provided a baseline to study MR basic concepts, its development steps and analytical models, a simulation toolkit, system types, and architectural types[5]. Egliston et al. propose how MR can be used in the future and its possibilities[6]. In the same way, Krishna et al. [7] and Flavian et al.[8] review how MR could be used in Smart Computing Education Systems and in customer experience. As MR is also the mix-ins of AR and VR, Cipresso et al., define the past, present, and future of VR and AR technology research[9]. Azuma et al. from Georgia Tech University, mention the recent advancement in Augmented Reality[10]. All these studies have focused on the applications of MR and its possibilities. Milgram [1] defined a 'continuum of Real to Virtual environments, where AR is one part of the general area of Mixed Reality'. In VR, the surrounding environment is virtual, while in AR the surrounding environment is real. There is a lack of sufficient study where MR could solve real-life problems and the challenges that should be faced which is presented in this research.

\section{Research Questions}
There are a few research questions that this paper aims to address for MR possibilities. This research addresses the following questions:
Q1. Can we see the increasing trend of applications of Mixed Reality in the future?
Q2. Could MR technologies be the interface of the future?
Q3. Are there potential challenges while exploring the applications of Mixed Reality?

\begin{figure}[hbt!]
  \centering
  \includegraphics[scale=0.60]{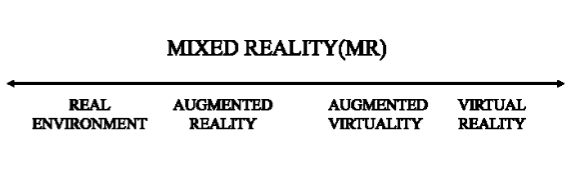}
  \caption{Reality-virtuality Continuum}
  \end{figure}

\section{Methodology}
To address the research question, specific methods are followed with different settings. Firstly, we reviewed previously published articles in MR-reputed journals and conferences. These articles were queried using specific keywords like "Mixed Reality", "Hololens", and "Extended Reality" in several databases and libraries such as IEEEXplore, Springer, ACM, Scopus, Science Direct, Microsoft Official Site, Google search, etc. The queried articles are related to Computer Engineering, Electrical Engineering, Computer Science, AR, VR, and MR domains. The papers that couldn't be accessed and required to pay are taken from the Sci-Hub website.
The published articles which are reviewed for this research range from the years 2000-2022. Many Mixed Reality articles were published within this period. And, within the span of the last five years, this technology is advancing rapidly. The research methodology is shown in Fig. 3 and consists of the following steps:

\begin{figure}[hbt!]
  \centering
  \includegraphics[scale=0.15]{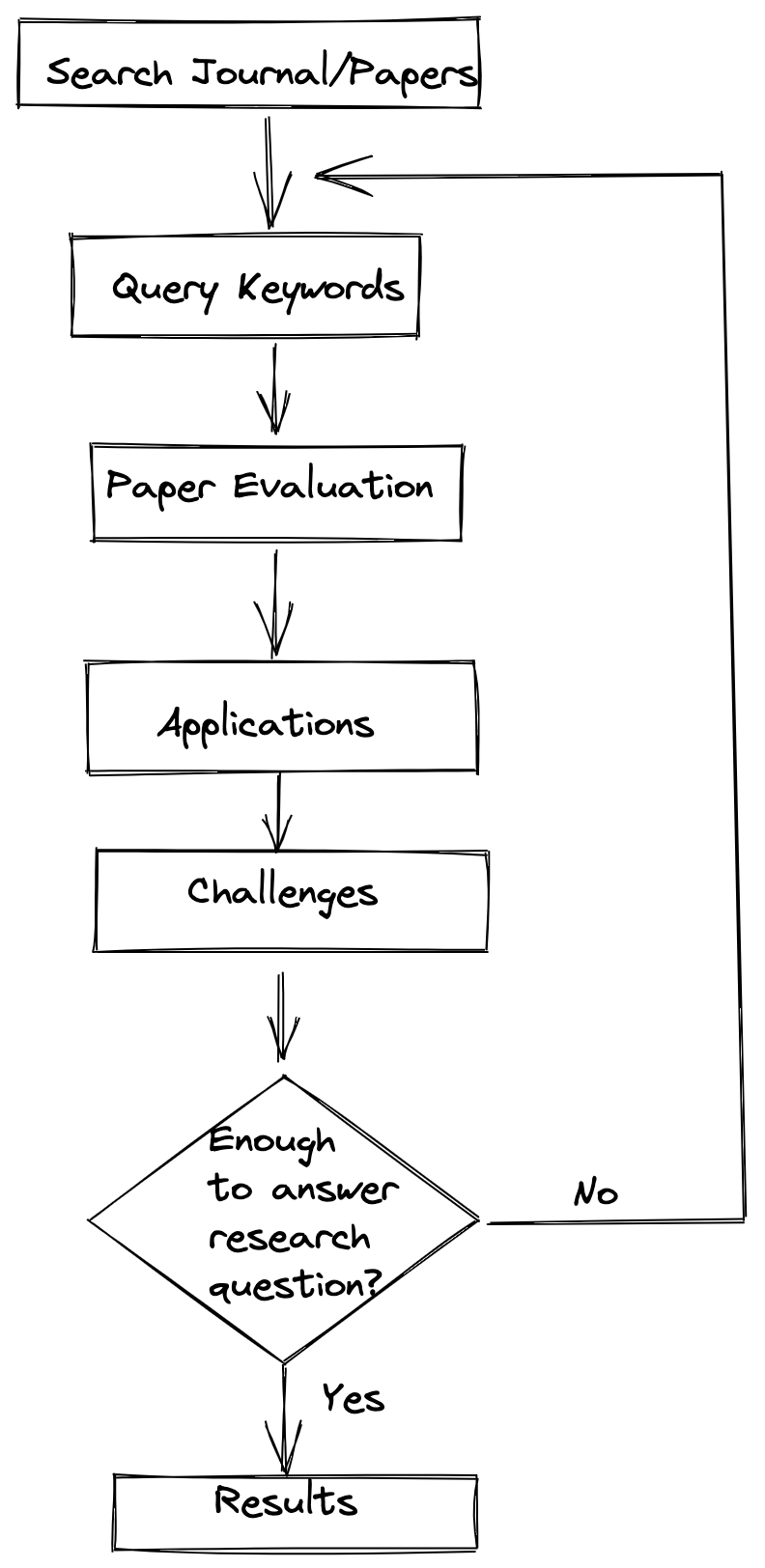}
  \caption{Research Methodology}
  \end{figure}
  
\begin{itemize}
    \item Articles were searched and reviewed using specific keywords related to mixed reality. Multiple databases were queried to find the relevant articles.
    \item Papers were evaluated qualitatively. The papers talk about different applications of mixed reality and how it is beneficial in the future.
    \item The applications areas are seen from medical, education, gaming, etc. that could solve the existing problem in the future. In addition to that, several challenges were scrutinized.
    \item More than 15 articles from general and specific fields were selected based on their relevance to the proposed research question. They are reviewed in detail within this research survey.
    \item The experiments and concepts regarding mixed reality are then presented sequentially which helps readers to understand easily.
\end{itemize}

\section{Experiments}

\subsection{Overview and Concept}
Mixed reality[3] is the technology that fuses the physical with the virtual world to provide upcoming evolution in human, computer, and environmental interactions. It's the next wave in computation. Before it, we had mainframes, PCs, and smartphones. Now, the concept of AR, VR, and MR are emerging which is mainstream for consumers and businesses. It liberates people from screen-bound experiences through interactions with data in the physical world. Mobile AR[11] nowadays offers mixed reality solutions, especially on social media. Generally, the applications to mixed reality also include environmental understanding i.e. anchors and spatial mapping, and human understanding i.e. eye-tracking, hand-tracking, and speech input, positioning, and locations in both physical and virtual spaces. 

To create true mixed reality interactions and experiences, there are three essential elements:
\begin{itemize}
    \item Input methods
    \item Perceptions from the surrounding
    \item Cloud-driven computer processing model
\end{itemize}

The reality-virtuality continuum[1] as shown in Fig. 2 represents the spectrum of realities as the mixed reality spectrum. At one end, we have the real physical system in which humans exist. On the other, we have the relative digital reality.

\subsection{Augmented and Virtual Reality}
Nowadays, many big tech companies are doing research on AR and VR. Snapchat uses AR features on image/video filters. However, most mobile devices and apps that exist on the market have little environmental perception capabilities. They cannot fully blend physical and digital realities. Augmented Reality[11] is achieved by overlaying holograms, graphics, images, and video streams in the physical world. Virtual reality has one difference in visualizing the real world. In VR, we cannot see the actual world which makes the experiences fully immersive digital experience. Mixed Reality can be experienced by transitioning between AR and VR which places a hologram in the physical world as if it is physically present. It represents the avatar digitally and personally present in the physical world in collaboration with people and objects.

\subsection{Mixed Reality Devices}
Mixed Reality perception is achieved by wearing a glass-mounted headset from which we can see the physical world to interact using digital assets/objects. The user can manipulate virtual objects to change the environment and interact using different sensors. For instance, a user can resize an object to fit into a room. There are several mixed reality devices available in the market in which hardware devices, chips, sensors, and software are integrated to perform a task in a coordinated way. MR devices can also act as standalone device, can be usable with glasses, and allows users to interact with their physical world. One limitation of MR headsets is it is still in the early stage of development which limits their production. Additionally, its field of view is often narrow. These advantages and limitations help to explore more possibilities in applications and challenges across Mixed Reality. Table I shows the different mixed reality devices being developed[12]. Fig 4. represents the decreasing price of VR over the years.

\begin{table}[htbp]
\caption{Mixed Reality Devices}
\begin{center}
\begin{tabular}{|c|c|c|c|c|}
\hline
\textbf{Brand} & \textbf{\textit{Product}}& \textbf{\textit{Release Date}}& \textbf{\textit{Country}}& \textbf{\textit{Price(US)}} \\
\hline
Zappar& ZapBox& 2018& UK& 30 \\
\hline
Tesseract& Holoboard& 2018& India& 349 \\
\hline
Occipital& Bridge& 2016& US& 399 \\
\hline
Nreal& Light& 2019& China& 499 \\
\hline
Dim. NXG& AjnaLens& 2018& India& 1500 \\
\hline
Magic Leap& One& 2018& US& 2295 \\
\hline
Microsoft& HoloLens& 2016& US& 3000 \\
\hline
Microsoft& HoloLens 2& 2019& US& 3500 \\
\hline
Avegant& LightField& 2016& US& N/A \\
\hline

\end{tabular}
\label{tab1}
\end{center}
\end{table}

\begin{figure}[hbt!]
  \centering
  \includegraphics[scale=0.15]{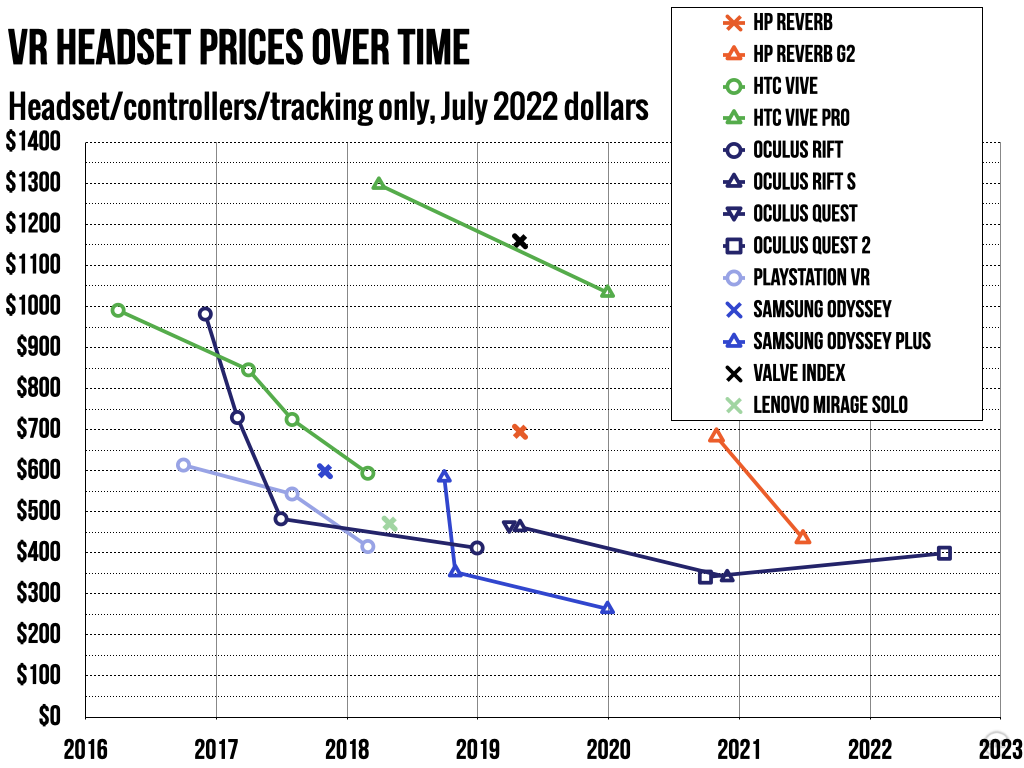}
  \caption{VR device prices each year}
  \end{figure}

\section{Applications}
Mixed Reality has a big potential to solve problems in various areas in the future. As this technology is growing, we can expect many possibilities and applications in different areas which are described below:

\subsection{Smart Education}
The integration of AR and MR technologies in mobile devices has opened up a lot of opportunities for applications in smart education. It overcomes the problems in the traditional way of teaching by showing its benefits. According to a recent study[7], the usage of mixed reality enhances students to learn complicated logical problems in Chemistry like a synthetic bond. They mentioned that mixed reality technologies help to picture the basic ideas of covalent bond to identify the objectives of chemical bonds. In the same way, a virtual learning environment builds interaction and cooperation among understudies and teachers. In the normal way of teaching, it's much more complicated to visualize higher-dimension data. Teachers can use Mixed Reality devices to simulate the mathematical analysis and graphs (higher dimensions) and teach their students so that it is easier for them to understand.

\begin{figure}[hbt!]
  \centering
  \includegraphics[scale=0.20]{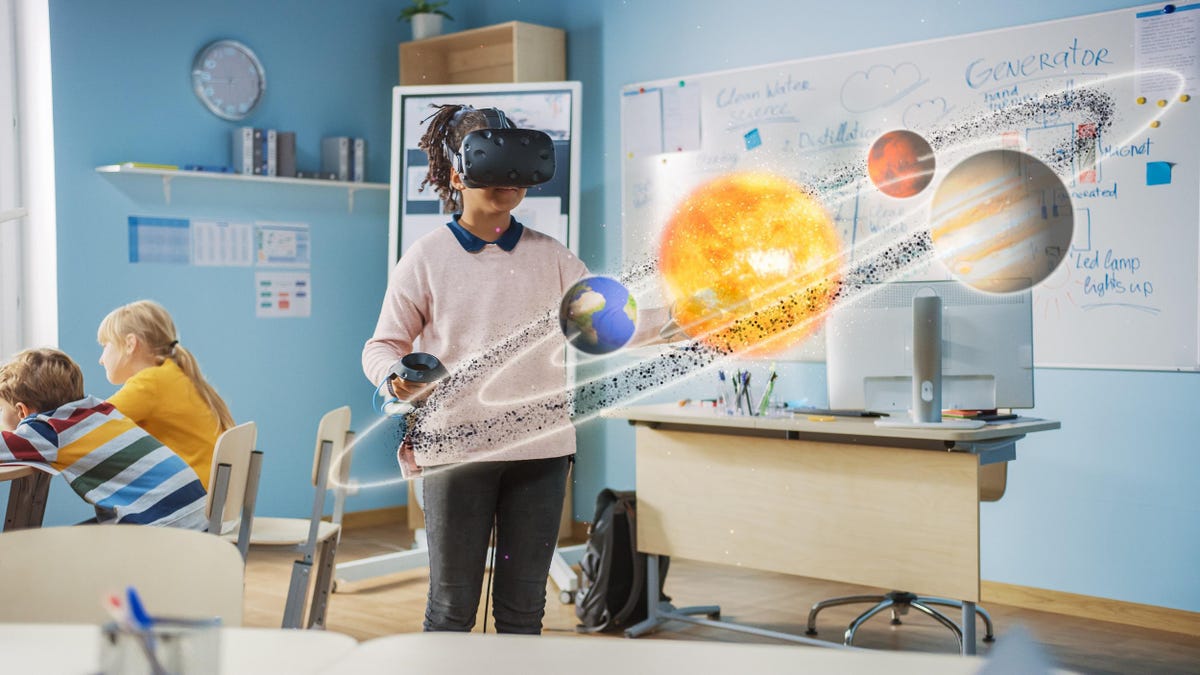}
  \caption{Mixed Reality in Education}
  \end{figure}

Pridhvi et al.[7] conducted the experiment to show the effectiveness of MR in the education system. They included different application development using Unity 3D and Unreal Engine to accomplish a practical look for learners/students. They visualized the Roman Colosseum in VR and also a low-resolution model of a maze. They also learned about the behavior of animals according to their adaptation to environments. S.V Viraktamath[14] and Touel et al.[13] their research mentions that MR devices can be used as a real-time analytical and interactive tool for instructors to promote the student's learning process. They concluded that the environment resulted in better prospects for education compared to the existing teaching methodology.

\subsection{Health System}
The study was conducted at the Norwegian University of Science and Technology and Gerup et al.[15] reveals that Mixed Reality is suitable for use in healthcare sectors because of its 3D capabilities and interactions, with scalable objects interacting simultaneously by various participants, and its ability to provide a deeper and better understanding of complex issues. In the same way, because of COVID-19 and social distancing, this technology has accelerated a lot.

 \begin{figure}[hbt!]
  \centering
  \includegraphics[scale=0.15]{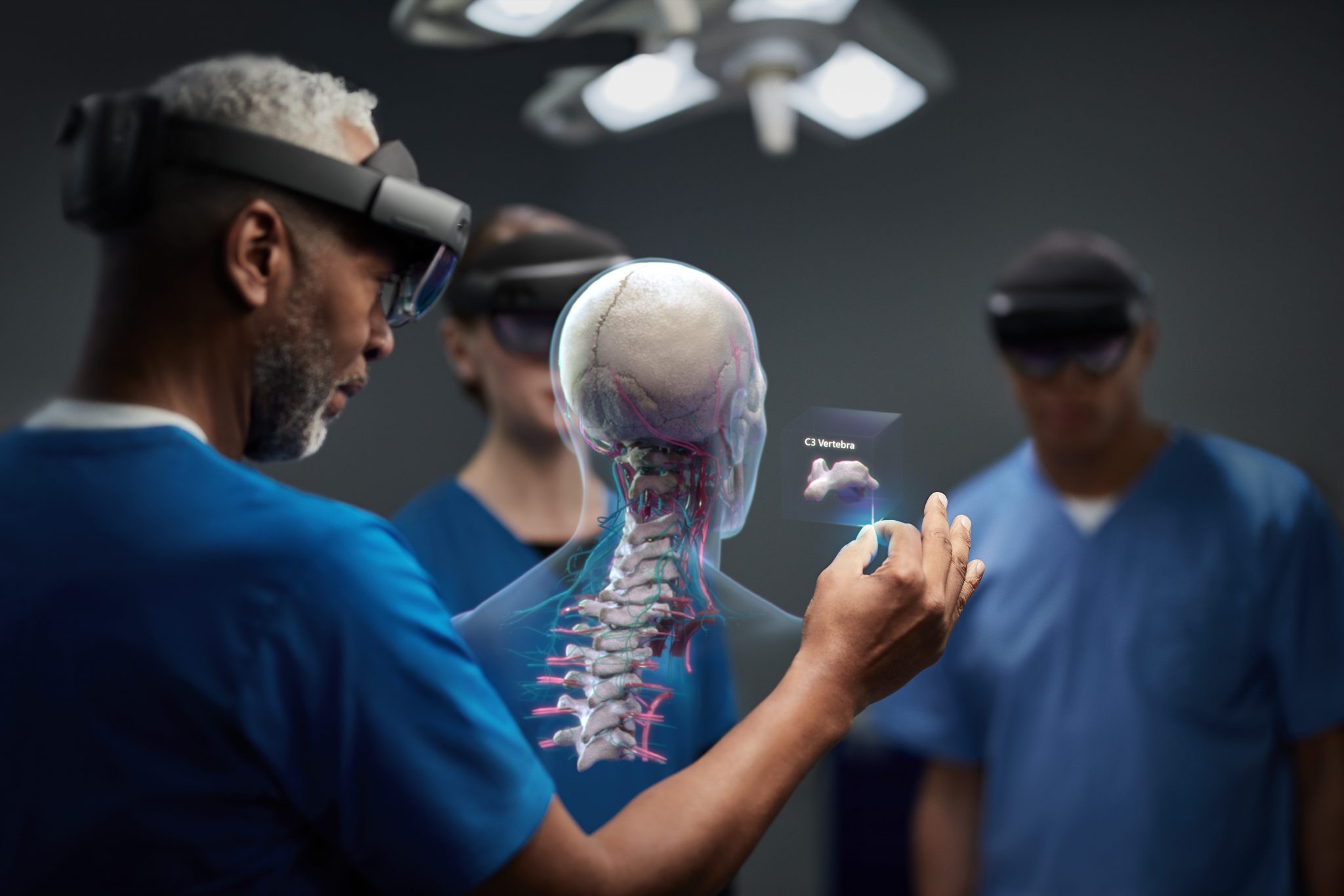}
  \caption{Mixed Reality in Health Care}
  \end{figure}
  
Similarly, Sánchez-Margallo JA[16] performed a study of the application of MR in medical training and surgical planning that focuses on minimally invasive surgery. They mentioned MR will reduce the time and effort while performing the training and surgery as medical training is a long and tiresome process. They conducted experiments on training in urology that allow visualization of 3D models, perform manipulation in different angles and dimensions, and able to interact independently with them. They found MR is a significant tool for studying and visualizing the human pelvic system and anatomy both individually and in groups.

\subsection{Engineering and Construction}
Construction projects contain a series of interdependent and interrelated activities. There are many hazardous works that could lead to serious accidents for the workers. Fei Dai et al. (2018) evaluated the feasibility of applying an emerging MR technology to enhance communication in the construction industry[17]. They developed a holographic application in Microsoft Hololens that helps in real-time collaboration in 3D space where the application was evaluated through trial and feedback from workers/participants in the construction industry. The experiment was conducted among 50 workers where they can communicate without being physically present at the site. Their experiments demonstrate that 58 percent of total responses agree MR has a positive impact on risk communication 34 percent are neutral. This helps to get a sense of how MR could be huge potential for applications in the construction industry.

 \begin{figure}[hbt!]
  \centering
  \includegraphics[scale=0.25]{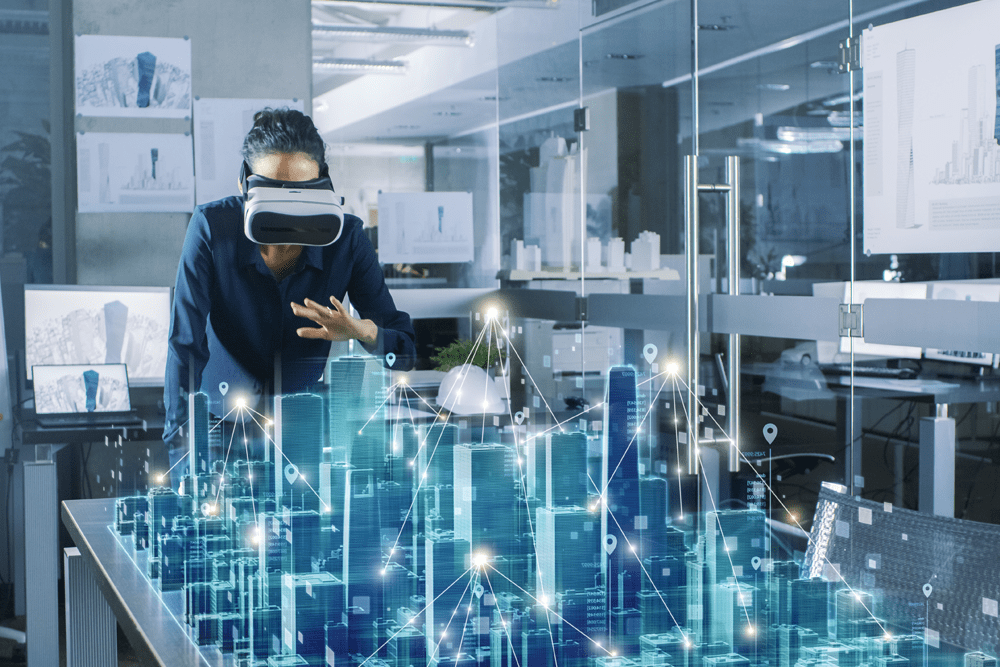}
  \caption{Mixed Reality in Engineering and Construction}
  \end{figure}

Similarly, MR can be used for engineering project planning to execution. For instance, MR can overlay certain elements onto a building plan so that investors/stakeholders can get a better understanding of the project. MR can be used to show 3D models, interact with them, provide tours, and track and document the progress of the project. In addition to this, they can be used for labeling and smooth collaboration, construction training, etc. These things seem to create a lot of potential in the construction industry using MR.

\subsection{Gaming and Entertainment}
Games and Entertainment are one of the major industries where Mixed Reality has been impacted the most. MR games use collaborative technologies to entertain, educate, motivate, and inspire people. The design of MR games and interactive gameplay environments has received a lot of attention across multiple disciplines. Elizabeth Bonsignore et al.[18] has mentioned that one example of gamification is the growth of context-aware apps like Foursquare which uses social and location-based services. Other games include iSpy and Parallel Kingdom which uses location-based services that offer a playful lens for visualizing experiences, and relationships of ideas grounded in real spaces i.e. crossing the boundary between real and virtual.

Likewise, many social media platform that includes interactive mixed-reality experiences are being developed. Roofie Du et al.[19] from Google presented Geollery, which is a social media platform to share, create, and explore geotagged information. It includes a real-time 'rendering of an interactive and collaborative mirrored world with 3D buildings, user-generated content, and geo-tagged social media. Using these features, users can view, chat, and collaborate with other participants remotely in the same spatial context in a virtual environment. This demonstrates MR features are being developed in the gaming and entertainment industries.

\subsection{Government Agency}
Mixed Reality also has applications across governmental bodies and agencies. A recent study conducted shows that AR and MR could be used for security in customs borders and protections. Similarly, they also mentioned envisioning transportation security 4.0 which reduces error and risk in security screening. In the same way, MR features can be used to study the probable natural calamities region and also have easy management to search and rescue.

\subsection{Retail and E-commerce}
MR can help to fill the lines between brick-and-mortar retail experiences and e-commerce. The products can be scanned using their tags and leverage important information from there so that people don't need to manually fill in the information. For instance, a retail mobile app can be used to identify all the products in a store automatically and add them to the shopping cart by adding computer vision and machine learning features.

\begin{figure}[hbt!]
  \centering
  \includegraphics[scale=0.15]{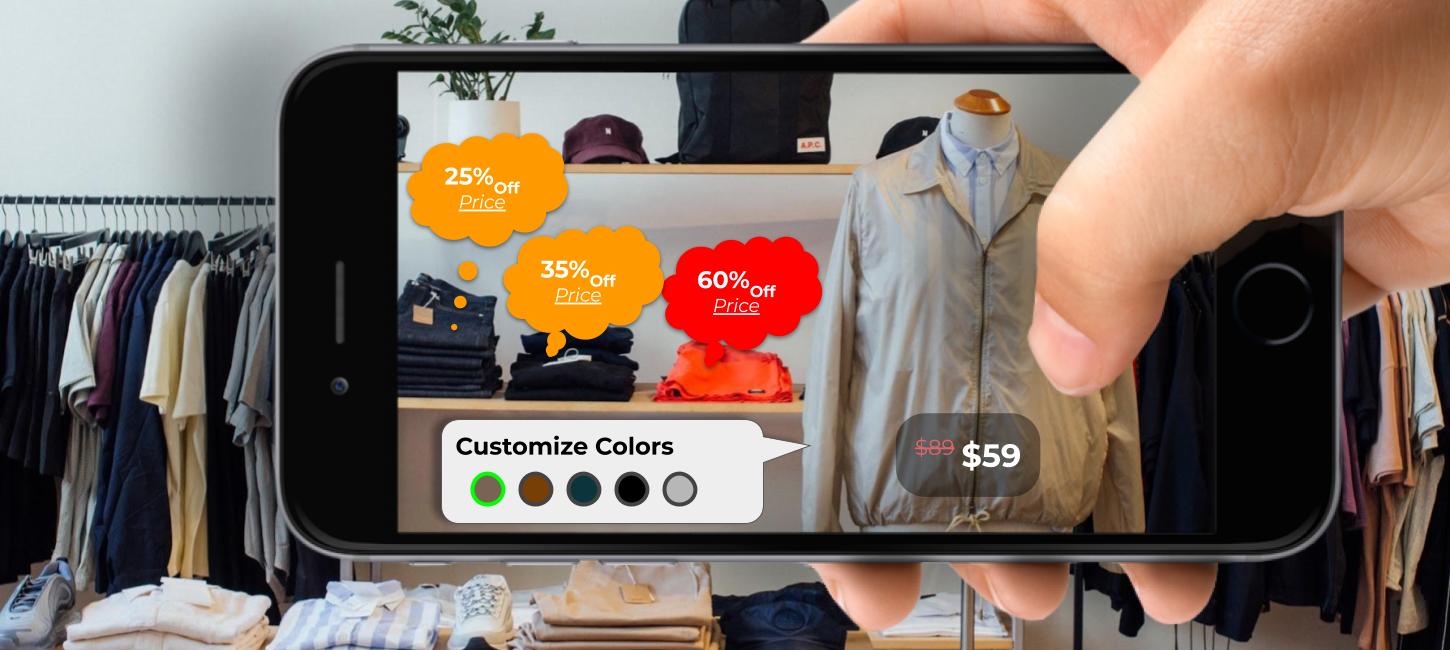}
  \caption{Mixed Reality in Retail and E-commerce}
  \end{figure}

\section{Challenges}
Although Mixed Reality induces a lot of potential benefits, there could be many challenges that need to be considered to operate it.
MR system development has a special software architecture that needs to execute MR algorithms faster in MR applications. It also has a similar architecture as others i.e. the server, client, and network. The server manages the data whereas the client is the user that executes the command. The network acts as a communicator between the server and the client. Some of the potential challenges of Mixed Reality lie in the following section:

\subsection{Architecture}
It's very important to design a suitable architecture to run MR applications smoothly. The architecture needs to support multiple users, who can simultaneously manipulate the objects in the environment. There should be non-noticeable latency in every operation so that user interaction would be smooth. As its architecture is based on client-server, there is a single point of failure. So, the server should be up and running every time for execution. This induces the challenges to be faced by developers in system design.

\subsection{Network Issue}
Networking is a most important part of MR that acts as a communicator between server and client. By distributing IP, every client can communicate with each other and access data through the server. If there is any delay in network communication, latency could be introduced which makes user interaction less convenient. Thus, the challenges also lie in the networking part.

\subsection{Data Loss and Security}
Data Loss and Security are one of the greatest challenges in Mixed Reality applications. The MR devices help to overlay the objects in the physical world where there is continuous data flow across the environment. The data could be lost on the way to overlaying the objects. In the meantime, anyone can interrupt the data flowing over the network which could create trouble. Moreover, the MR applications used in the government sector need to be more secure in terms of data handling and processing.

\subsection{Pedagogical Challenges}
When MR is used in educational settings, it changes past pedagogical techniques. The study by Glover and Miller[20] in 2001 mentions that it is a prerequisite to accepting technology by teachers and students. Both teachers and students should be prepared for such technologies. Similarly, learning activities should include well-designed interactions with friends and the system. There should be minimum information overload and field-of-view should be accurate. Otherwise, the visualizations are perceived as obtrusive.

\subsection{Technical Challenges}
MR technology and devices are still in their infancy stage. Pascal Knierim et al.[21] mentioned that it includes a small view field, limited battery life, uncomfortable wearing devices for a long time, etc. From the fast technological advances across MR, these challenges could be solved in the future. Currently, many advanced features like object recognition, hand gesture recognition, and speech identification are being integrated into MR. But it's still a research question of how to design an interactive concept for MR user experiences.

\section{Results}

We reviewed several research papers in the field of Mixed Reality and studied the potential benefits and challenges. From the study we conducted, the research question that is presented could be answered.

\subsection{Can we see the increasing trend of applications of Mixed Reality in the future?}
\begin{itemize}
    \item The experiments that are performed to analyze the applications of MR in different fields above show that MR has the potential to be used for solving the existing problem and a lot of benefits can be leveraged from it. It has increasing benefits across education, health, construction, gaming, etc. which could be an interface of the future.
    \item Despite some negative impacts that MR could bear, we could see a lot of positive impacts of MR in human lives. This answers our first research question.
\end{itemize}

\subsection{Could MR technologies be the interface of a future?}  
\begin{itemize}
    \item As seen from the increasing number of companies and MR devices year to year in Table I, a lot of big companies are spending billions to adopt MR technologies as shown in Table I. Also, the applications seen in various sectors have also shaped MR to be one of the emerging technologies. This demonstrates MR technology as the interface of the future.
\end{itemize}

\subsection{Are there potential challenges while exploring the applications of Mixed Reality?}
\begin{itemize}
    \item Certain challenges to adopting MR technologies in various fields are studied above. In the same way, there are some uses of MR that can impact human lives negatively. These all challenges should be cautiously handled in order to fully utilize the benefits of Mixed Reality. These aspects answer that there are certain challenges to exploiting the use of Mixed Reality.
\end{itemize}

\section{Conclusion}
Mixed Reality is an emerging technology with a lot of potential that is being developed and researched across the globe. Multi-national companies are spending huge amounts to explore the applications of MR so that it would solve human problems in the future. This research helps to study the potential benefits and challenges of MR in the future and the way MR could be an interface to the future. There are still many areas like sports and transportation that are not explored in this survey. Also, MR could impact negatively in some areas which is not described in this research. Overall, MR technology has a potential role in solving human life problems.

\vspace{12pt}

\end{document}